\documentclass[prl,showpacs,twocolumn,floatfix]{revtex4}%
\usepackage{amsmath}
\usepackage{graphicx}
\usepackage{amsfonts}
\usepackage{amssymb}%

\newcommand{\ket}[1]{\ensuremath{\left\vert #1 \right\rangle}}

\newcommand{\boldvec}[1]{\ensuremath{\mbox{\boldmath${\mathbf{#1}}$}}}
\newcommand{\spvec}[1]{\ensuremath{\boldvec{#1}\,}}
\newcommand{\vecop}[1]{\ensuremath{\boldvec{\hat{#1}}}}

\newcommand{\unitvec}[1]{\ensuremath{\hat{#1}}}
\newcommand{\levellabel}[1]{\ensuremath{\ket{#1}}}
\newcommand{\isolevind}[2]{\ensuremath{#1^{(#2)}}}
\newcommand{\isolevlab}[2]{\ensuremath{\levellabel{\isolevind{#1}{#2}}}}
\newcommand{\cgcoeff}[6]{\ensuremath{C^{#1\; #2\; #3}_{#4\; #5\; #6}}}

\begin{document}
\title{Dual species matter qubit entangled with light}
\date{\today }
\author{S.-Y. Lan, S. D. Jenkins,$^{*}$ T. Chaneli\`{e}re,$^{\star}$ D. N. Matsukevich,$^{\dagger}$
C. J. Campbell, R. Zhao, T. A. B. Kennedy, and A. Kuzmich}
\affiliation{School of Physics, Georgia Institute of Technology,
Atlanta, Georgia 30332-0430} \pacs{42.50.Dv,03.65.Ud,03.67.Mn}
\begin{abstract}
  We propose and demonstrate an atomic
  qubit based on a cold $^{85}$Rb-$^{87}$Rb isotopic mixture, entangled with a frequency-encoded optical qubit.
The interface of an atomic qubit with a single spatial light mode,
and the ability to independently address the two atomic qubit
  states, should provide the basic element of an
  interferometrically robust quantum network.
\end{abstract}
\maketitle

Quantum mechanics permits the secure communication of information
between remote parties \cite{bennett,ekert,bouwmeester}. However,
direct optical fiber based quantum communication over distances
greater than about 100 km is challenging due to intrinsic fiber
losses. To overcome this limitation it is necessary to take
advantage of quantum state storage at intermediate locations on the
transmission channel. Interconversion of the information from light
to matter to light is therefore essential. It was the necessity to
interface photonic communication channels and storage elements that
lead to the proposal of the quantum repeater as an architecture for
long-distance distribution of quantum information via qubits
\cite{briegel,duan}.

Recently there has been rapid progress in interfacing photonic and
stored atomic qubits. Two-ensemble encoding of matter qubits was
used to achieve entanglement of photonic and atomic rubidium qubits
and quantum state transfer from matter to light \cite{matsukevich}.
This was followed by a more robust single-ensemble qubit encoding
\cite{matsukevich1}, which led to full light-matter-light qubit
interconversion and entanglement of two remote atomic qubits
\cite{matsukevich2}. More recently, both two-ensemble and
single-ensemble atomic qubits were reported using cesium gas
\cite{chou,riedma}.

To realize scalable long distance qubit distribution
telecommunication-wavelength photons and long-lived quantum memory
elements are required \cite{chaneliere}. Although multiplexing of
atomic memory elements vastly improves the dependence of
entanglement distribution on storage lifetime \cite{collins}, there
remains, the problem of robust atomic and photonic qubits for
long-distance communication. Two-ensemble encoding suffers from the
problem of long-term interferometric phase stability, while qubit
states encoded in a single ensemble are hard to individually
address.

\begin{figure}
  \centering
  \vspace{-0.0cm}
  \includegraphics[width=3.0in]{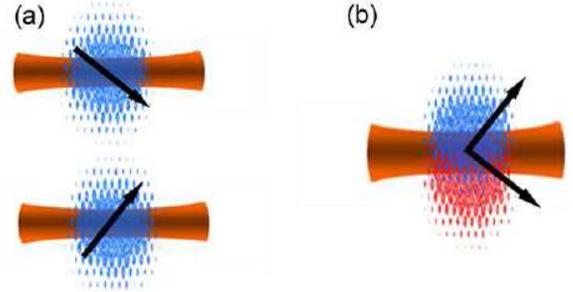}
 \vspace{-0.2cm}
  \caption{(a) Schematic shows two orthogonal qubit states (arrows) encoded in
two atomic ensembles coupled to distinct spatial light modes
\cite{duan,matsukevich}, (b) the proposed architecture encodes the
states in a two species atomic mixture, coupled to a single light
mode.}
  \label{fig:1}
\end{figure}

A protocol for implementing entanglement distribution with an atomic
ensemble-based quantum repeater has been proposed \cite{duan}. It
involves generating and transmitting each of the qubit basis states
individually, in practice via two interferometrically separate
channels. Under prevailing conditions of low overall efficiencies it
provides improved scaling compared to direct qubit entanglement
distribution \cite{briegel}. Its disadvantage is the necessity to
stabilize the length of both transmission channels to a small
fraction of the optical wavelength, as the distribution of qubit
entanglement is sensitive to the relative phase fluctuations in the
two arms.

In this Letter we propose an interferometrically robust quantum
repeater element based on entangled mixed species atomic, and
frequency-encoded photonic, qubits, Fig. 1. This avoids the use of
two interferometrically separate paths for qubit entanglement
distribution. The qubit basis states are encoded as single spin wave
excitations in each one of the two atomic species co-trapped in the
same region of space. The spectroscopically resolved transitions
enable individual addressing of the atomic species. Hence one may
perform independent manipulations in the two repeater arms which
share a single mode transmission channel. Phase stability is
achieved by eliminating the relative ground state energy shifts of
the co-trapped atomic species, as is in any case essential to
successfully read out an atomic excitation \cite{revival}.

\begin{figure}
  \centering
  \vspace{-0.15cm}
  \includegraphics[width=3.0in]{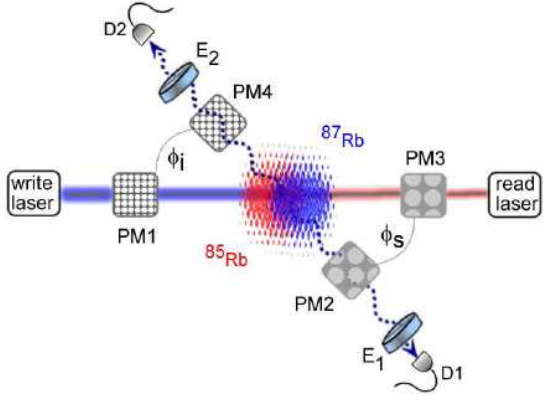}
  \vspace{-0.1cm}
  \caption{Schematic of the experimental set-up showing the geometry
    of the addressing and scattered fields from the co-trapped isotope
    mixture of $^{85}$Rb-$^{87}$Rb. The write and read laser fields
    generate signal and idler fields, respectively detected at D1 and D2; E$_1$, E$_2$ are optical
    frequency filters. PM1-4 are light phase modulators, $\phi_s$ and $\phi_i$ are relative phases of the driving rf fields,
    see text for details.}
  \label{fig:Schematic}
\end{figure}

We consider a co-trapped isotope mixture of $^{85}$Rb and $^{87}$Rb,
containing, respectively, $N_{85}$ and $N_{87}$ atoms cooled in a
magneto optical trap, as shown in Fig. 2. Unpolarized atoms of
isotope $\nu$ ($\nu \in \{85, 87\}$) are prepared in the ground
hyperfine level $\isolevlab{a}{\nu }$, where $\isolevlab{a}{85}
\equiv \levellabel{5S_{1/2},F_a^{(85)} = 3}$, $\isolevlab{a}{87}
\equiv \levellabel{5S_{1/2},F_a^{(87)} = 2}$, and $F_f^{(\nu )}$ is
the total atomic angular momentum for level $\isolevlab{f}{\nu }$.
We consider the Raman configuration with ground levels
$\isolevlab{a}{\nu}$ and $\isolevlab{b}{\nu}$ and excited level
$\isolevlab{c}{\nu}$ with energies $\hbar\omega_{a}^{(\nu )}$,
$\hbar\omega_{b}^{(\nu )}$, and $\hbar\omega_{c}^{(\nu )}$
respectively. Level $\isolevlab{b}{\nu }$ corresponds to the ground
hyperfine level with smaller angular momentum, while level
$\isolevlab{c}{\nu }$ is the $\levellabel{5P_{1/2}}$ hyperfine level
with $F_c^{(\nu )}=F_a^{(\nu )}$. A 150 ns long \textit{write} laser
pulse of wave vector $\mathbf{k}_w = k_w\unitvec{y}$, horizontal
polarization $\mathbf{e}_H = \unitvec{z}$ and temporal profile
$\varphi(t)$ (normalized to unity $\int dt~|\varphi(t)|^2 =1$)
impinges on an electro-optic modulator (EOM), producing sidebands
with frequencies $ck_w^{(85)} = ck_w + \delta \omega_w$ and
$ck_w^{(87)} = ck_w - \delta\omega_w$ ($\delta \omega_w = 531.5$
MHz) nearly resonant on the respective isotopic $D_1$
($\isolevlab{a}{\nu } \leftrightarrow \isolevlab{c}{\nu }$)
transitions with detunings $\Delta_{\nu } = ck_w^{(\nu )} -
(\omega_c^{(\nu )} - \omega_a^{(\nu )}) \approx -10 $ MHz.
Spontaneous Raman scattering of the \textit{write} fields results in
signal photons with frequencies $ck_s^{(\nu )} = ck_w^{(\nu )} +
(\omega_b^{(\nu )} - \omega_a^{(\nu )})$ on the $\isolevlab{b}{\nu }
\leftrightarrow \isolevlab{c}{\nu }$ transitions. The positive
frequency component of the detected signal electric field from
isotope $\nu $ with vertical polarization $\mathbf{e}_V$ is given by
\begin{eqnarray}
  \vecop{E}^{(\nu)(+)}(\mathbf{r},t) &=& \sqrt{\frac{\hbar
      k_s^{(\nu)}}{2\epsilon_0}}
  e^{-ick_s^{(\nu)} (t - \unitvec{k}_s^{(\nu)} \cdot \mathbf{r})}
  \nonumber \\
  & & \times u_s(\mathbf{r}) \hat{\psi}_s^{(\nu)}(t-\unitvec{k}_s\cdot
  \mathbf{r}) \mathbf{e}_V \mbox{,}
  \label{eq:SigEfield}
\end{eqnarray}
where $u_s(\spvec{r})$ is the transverse spatial profile of the
signal field (normalized to unity in its transverse plane), and
$\hat{\psi}_s^{(\nu)}(t)$ is the annihilation operator for the
signal field. These operators obey the usual free field, narrow
bandwidth bosonic commutation relations $[ \hat{\psi}_s^{(\nu)}(t),
\hat{\psi}_s^{(\nu')\dag}(t') ] = \delta_{\nu,\nu'} \delta(t-t')$.
The emission of $V$-polarized signal photons creates correlated
atomic spin-wave excitations with annihilation operators given by
\begin{equation}
  \hat{s}^{(\nu)} = \cos\theta_\nu \hat{s}_{-1}^{(\nu)} -
  \sin\theta_\nu \hat{s}_{+1}^{(\nu)} \mbox{,}
  \label{eq:HspinWaveDef}
\end{equation}
where $$\cos^2\theta_\nu = \sum_{m=-F_a^{(\nu)}}^{F_a^{(\nu)}}
X_{m,-1}^{(\nu)\,2} / \sum_{\alpha = \pm1}
\sum_{m=-F_a^{(\nu)}}^{F_a^{(\nu)}} X_{m,\alpha}^{(\nu)\,2},$$
$X_{m,\alpha}^{(\nu)} \equiv
\cgcoeff{F_a^{(\nu)}}{1}{F_c^{(\nu)}}{m}{0}{m}
\cgcoeff{F_b^{(\nu)}}{1}{F_c^{(\nu)}}{m-\alpha}{\alpha}{m} $ is a
product of Clebsch-Gordan coefficients, and the spherical vector
components of the spin wave are given by
\begin{equation}
  \hat{s}_\alpha^{(\nu)} = \sum_{m=-F_a^{(\nu)}}^{F_a^{(\nu)}}
  \frac{X_{m,\alpha}^{(\nu)}}{\sqrt{\sum_{m=-F_a^{(\nu)}}^{F_a^{(\nu)}}
      \left|X_{m,\alpha}^{(\nu)}\right|^2}}
  \hat{s}_{m,\alpha}^{(\nu)} \mbox{.}
  \label{eq:spSwaveComp}
\end{equation}
The spin wave Zeeman components of isotope $\nu$ are given in terms
of the $\mu$-th $^\nu$Rb atom transition operators
${\sigma}_{\isolevind{a}{\nu},m;\: \isolevind{b}{\nu},m'}$ and the
\textit{write} $u_w(\spvec{r})$ and signal $u_s(\spvec{r})$ field
spatial profiles
\begin{eqnarray}
  \hat{s}_{m,\alpha}^{(\nu)} &=&
  i{A}^{(\nu)}\sqrt{  \frac{(2F_a^{(\nu )}+1)}{N_\nu} }
  \sum_{\mu}^{N_\nu}
  {\sigma}_{\isolevind{a}{\nu},m;\: \isolevind{b}{\nu},
    m}^{\mu} \nonumber\\
  & &\times e^{i\left(\mathbf{k}_s^{(\nu)}-\mathbf{k}_w^{(\nu)}\right) \cdot
    \mathbf{r}_{\mu}} u_s(\mathbf{r}_{\mu})
  u_w^{\ast}(\mathbf{r}_{\mu}) .
  \label{eq:swavemCompDef}
\end{eqnarray}
The effective overlap of the \textit{write} beam and the detected
signal mode \cite{jenkins} is given by
\begin{equation}
  {A}^{(\nu)} = \left(\int d^3r \left| u_s(\spvec{r})
      u_w^{\ast}(\spvec{r}) \right|^2
    \frac{n^{(\nu)}(\spvec{r})}{N_\nu} \right)^{-1/2},
  \label{eq:AbarDef}
\end{equation}
where $n^{(\nu)}(\spvec{r})$ is the number density of isotope $\nu$.
The interaction responsible for scattering into the collected signal
mode is given by
\begin{eqnarray}
  \hat{H}_s(t) &=& i\hbar\chi\varphi(t) \Bigl( \cos\eta\,
    \hat{\psi}_s^{(85)\dag}(t) \hat{s}^{(85)\dag} \nonumber\\
    & & + \sin\eta\,
    \hat{\psi}_s^{(87)\dag}(t) \hat{s}^{(87)\dag}  \Bigr) + h.c. \mbox{,}
  \label{eq:sigHam}
\end{eqnarray}
where $\chi \equiv \sqrt{\chi_{85}^2 + \chi_{87}^2}$ is a
dimensionless interaction parameter,
\begin{equation}
  \chi_\nu \equiv \frac{\sqrt{2}d_{cb}^{(\nu)}d_{ca}^{(\nu)}}{{A}^{(\nu)}\Delta_\nu}
  \frac{k_s^{(\nu)}k_w^{(\nu)}n_w^{(\nu)} N_\nu}
  {(2F_a^{(\nu )}+1)\hbar \epsilon_0 } \sqrt{\sum_{\alpha=\pm 1} \sum_{m=-F_a^{(\nu)}}^{F_a^{(\nu)}}
    \left|X_{m,\alpha}^{(\nu)}\right|^2} \mbox{,}
  \label{eq:chi_isoDef}
\end{equation}
$d_{ca}^{(\nu)} $ and $d_{cb}^{(\nu)} $ are reduced matrix elements,
$n_w^{(\nu)}$ is the average number of photons in the \textit{write}
pulse sideband with frequency $ck_w^{(\nu)}$, and the parametric
mixing angle $\eta$ is given by $ \cos^2\eta =
\chi_{85}^2/(\chi_{85}^2 + \chi_{87}^2)$. The interaction picture
Hamiltonian also includes terms representing Rayleigh scattering and
Raman scattering into undetected modes. One can show, however, that
these terms commute with the signal Hamiltonian
(Eq.~(\ref{eq:sigHam})) and with the operators
$\hat{\psi}_s^{(\nu)}(t)$ and $\hat{s}^{(\nu)}$ to order
$O(1/\sqrt{N})$. As a result, the interaction picture density
operator for the signal-spin wave system (tracing over undetected
field modes) is given by $\hat{U}\hat{\rho}_0\hat{U}^{\dag}$, where
${\hat{\rho}_0}$ is the initial density matrix of the unpolarized
ensemble and the vacuum electromagnetic field, and the unitary
operator $\hat{U}$ is given by
\begin{equation}
  \ln \hat{U} = {\chi (\cos\eta \hat{a}^{(85)\dag}\hat{s}^{(85)\dag}
      + \sin\eta \hat{a}^{(87)\dag}\hat{s}^{(87)\dag} -
      h.c. )},
  \label{eq:Udef}
\end{equation}
where  $\hat{a}^{(\nu)} = \int dt \varphi^{\ast}(t)
\hat{\psi}_s^{(\nu)}(t)$ is the discrete signal mode bosonic
operator. When the \textit{write} pulse is sufficiently weak we may
write $\hat{U}-1 = \chi (\cos\eta
\hat{a}^{(85)\dag}\hat{s}^{(85)\dag} + \sin\eta
\hat{a}^{(87)\dag}\hat{s}^{(87)\dag}) + O(\chi^2)$, i.e., the Raman
scattering produces entanglement between a two-mode field (frequency
qubit) and the isotopic spin wave (dual species matter qubit).
Although we explicitly treat isotopically distinct species, it is
clear that the analysis is easily generalized to chemically distinct
atoms and/or molecules.

To characterize the nonclassical correlations of this system, the
signal field is sent to an electro-optic phase modulator (PM2 in
Fig. 2) driven at a frequency $\delta\omega_s = \delta\omega_w -
\big[ \big(\omega_a^{(87)} -\omega_b^{(87)}\big) -
\big(\omega_a^{(85)} - \omega_b^{(85)}\big) \big] /2 = 1368$ MHz.
The modulator combines the two signal frequency components into a
central frequency $ck_s = c(k_s^{(85)} + k_s^{(87)})/2$ with a
relative phase $\phi_s$. A photoelectric detector preceded by a
filter (an optical cavity, E1 in Fig. 2) which reflects all but the
central signal frequency is used to measure the statistics of the
signal. We describe the detected signal field using the bosonic
field operator,
\begin{eqnarray}
\hat{\psi}_s(t,\phi_s)=\sqrt{\frac{\epsilon_s^{(85)}}{2}}
e^{-i\phi_s/2} \hat{\psi}_{s}^{(85)}(t) +
\sqrt{\frac{\epsilon_s^{(87)}}{2}}e^{i\phi_s/2}
    \hat{\psi}_s^{(87)}(t)  \nonumber\\
+\sqrt{\frac{1-\epsilon_s^{(85)}}{2}}e^{-i\phi_s/2}
\hat{\xi}_s^{(85)}(t)+\sqrt{\frac{1-\epsilon_s^{(87)}}{2}}e^{i\phi_s/2}
\hat{\xi}_s^{(87)}(t) \nonumber
  \label{eq:detSigFOp}
\end{eqnarray}
where $\epsilon_s^{(\nu)}  \in [0,1]$ is the signal efficiency
including propagation losses and losses to other frequency sidebands
within PM2, and $\hat{\xi}_s^{(\nu)}(t)$ represents concomitant
vacuum noise. While quantum memory times in excess of 30 $\mu$s have
been demonstrated \cite{dspg}, here the spin wave qubit is retrieved
after 150 ns by shining a vertically polarized \textit{read} pulse
into a third electro-optic phase modulator (PM3 in Fig. 2),
producing two sidebands with frequencies $ck_r^{(85)}$ and
$ck_r^{(87)}$ resonant on the $\isolevlab{b}{85} \leftrightarrow
\isolevlab{c}{85}$ and $\isolevlab{b}{87} \leftrightarrow
\isolevlab{c}{87}$ transitions, respectively. This results in the
transfer of the spin wave excitations to horizontally polarized
idler photons emitted in the phase matched directions
$\spvec{k}_i^{(\nu)} = \spvec{k}_w^{(\nu)} - \spvec{k}_s^{(\nu)} +
\spvec{k}_r^{(\nu)}$. We treat the retrieval dynamics using the
effective beam splitter relations $\hat{b}^{(\nu)} =
\sqrt{\epsilon_r^{(\nu)}} \hat{s}^{(\nu)} +
\sqrt{1-\epsilon_r^{(\nu)}} \hat{\xi}_r^{(\nu)}$, where
$\epsilon_r^{(\nu)}$ is the retrieval efficiency of the spin wave
stored in the isotope $^\nu$Rb, $\hat{b}^{(\nu)} = \int dt
\varphi_i^{(\nu)\ast}(t) \hat{\psi}_i^{(\nu)}(t)$ is the discrete
idler bosonic operator for an idler photon of frequency
$ck_i^{(\nu)}$, $\varphi_i^{(\nu)}(t)$ is the temporal profile of an
idler photon emitted from the $^\nu$Rb spin wave (normalized to
unity), and $\hat{\psi}_i^{(\nu)}(t)$ is the annihilation operator
for an idler photon emitted at time $t$. As with the signal
operators, the idler field operators obey the usual free field,
narrow bandwidth bosonic commutation relations
$[\hat{\psi}_i^{(\nu)}(t), \hat{\psi}_i^{(\nu')\dag}(t')] =
\delta_{\nu,\nu'} \delta(t-t')$. A fourth EOM, PM4, driven at a
frequency $\delta\omega_i = \delta\omega_w - (\Delta_{85} +
\Delta_{87})/2=531.5$ MHz combines the idler frequency components
into a sideband with frequency $ck_i = c(k_i^{(85)} + k_i^{(87)})/2$
with a relative phase $\phi_i$. The combined idler field is measured
by a photon counter preceded by a frequency filter (an optical
cavity, E2 in Fig. 2) which only transmits fields of the central
frequency $ck_i$. The detected idler field is described by the
bosonic field operator,
\begin{eqnarray}
\hat{\psi}_i(t,\phi_i) = \sqrt{\frac{\epsilon_i^{(85)}}{2}}
  e^{i\phi_i/2} \hat{\psi}_{i}^{(85)}(t) + \sqrt{\frac{\epsilon_i^{(87)}}{2}}e^{-i\phi_i/2}
    \hat{\psi}_i^{(87)}(t)  \nonumber\\
+\sqrt{\frac{1-\epsilon_i^{(85)}}{2}}e^{i\phi_i/2}
\hat{\xi}_i^{(85)}(t)+\sqrt{\frac{1-\epsilon_i^{(87)}}{2}}e^{-i\phi_i/2}
\hat{\xi}_i^{(87)}(t)  \nonumber
\end{eqnarray}
where $\epsilon_i^{(\nu)} \in [0,1]$ is the idler efficiency
including propagation losses and losses to other frequency sidebands
within PM4, and $\hat{\xi}_i^{(\nu)}(t)$ represents associated
vacuum noise. The {\it write-read} protocol in our experiment is
repeated $2\cdot 10^5$ times per second.

The signal-idler correlations result in phase-dependent coincidence
rates given, up to detection efficiency factors, by
$C_{si}(\phi_s,\phi_i) = \int dt_s \int dt_i \left\langle
  \hat{\psi}_s^{\dag}(t_s,\phi_s)\hat{\psi}_i^{\dag}(t_i,\phi_i)
  \hat{\psi}_i(t_i,\phi_i) \hat{\psi}_s(t_s,\phi_s) \right\rangle$.
From the state of the atom-signal system after the \textit{write}
process, $\hat{U}\hat{\rho}_0\hat{U}^{\dag}$, (Eq.\ref{eq:Udef}), we
calculate the coincidence rates to second order in $\chi$,
\begin{eqnarray}
  \lefteqn{C_{si}(\phi_s,\phi_i) = \frac{\chi^2 }{4}
    \biggl(  \mu^{(85)}\cos^2\eta + \mu^{(87)} \sin^2\eta}
    \nonumber\\
    & &+ \Upsilon \sqrt{\mu^{(85)}\mu^{(87)}} \sin 2\eta
     \cos\left(\phi_i-\phi_s+\phi_0\right) \biggr)
  \label{eq:fringes}
\end{eqnarray}
where $\mu^{(\nu)}\equiv \epsilon_r^{(\nu)}
\epsilon_i^{(\nu)}\epsilon_s^{(\nu)}$, and $\Upsilon$ and $\phi_0$
represent a real amplitude and phase, respectively, such that
\begin{equation}
  \Upsilon e^{-i\phi_0} = e^{-\left(\delta\phi_s^2 +
      \delta\phi_i^2\right)/2} \int dt
    \varphi_i^{(85)*}(t)\varphi_i^{(87)}(t) \mbox{,}
  \label{eq:upsphi0Def}
\end{equation}
and we account for classical phase noise in the rf driving of the
EOM pairs PM1,4 and PM2,3, by treating $\phi_s$ and $\phi_i$ as
Gaussian random variables with variances $\delta\phi_s^2$ and
$\delta\phi_i^2$ respectively, see Fig. 2. When the write fields are
detuned such that the rates of correlated signal-idler coincidences
are equal (i.e., when $\mu^{(85)} \cos^2\eta = \mu^{(87)}
\sin^2\eta$), the fringe visibility is maximized, and Eq.
(\ref{eq:fringes}) reduces to
\begin{equation}
  C_{si}(\phi_s,\phi_i)=\frac{\chi^2}{2} \mu^{(85)}\cos^2\eta
[ 1 + \Upsilon \cos (\phi_i-\phi_s+\phi_0) ].
  \label{eq:fringes1}
\end{equation}
\begin{figure}
  \centering
  \includegraphics[width=2.7in]{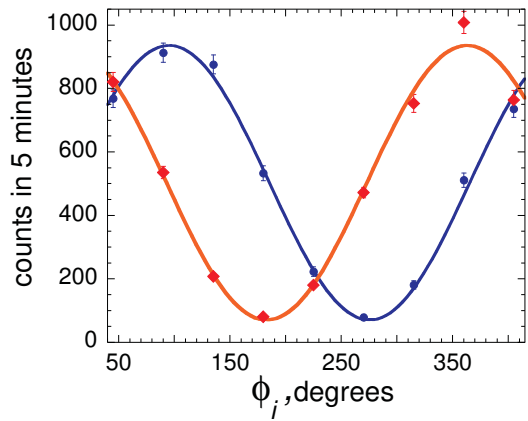}
  \vspace{-0.2cm}
  \caption{Measured $C_{si}(\phi_s,\phi_i)$ as a function
    of $\phi_i$ for $\phi_s=0$, diamonds and for
    $\phi_s=-\pi/2$, circles. The angle $\phi_0$ is absorbed into the arbitrary definition of the origin, i.e., $\phi_0$ is defined to be zero.
    Solid lines are sinusoidal
    fringes based on Eq. (\ref{eq:fringes1}) with $\Upsilon = 0.86$.
    Single channel counts of D1 and D2 show no
    dependence on the phases.}
  \label{fig:cFringes}
\end{figure}

Fig. 3 shows coincidence fringes as a function of $\phi_i$ taken for
two different values of $\phi_s$. The detection rates measured
separately for $^{85}$Rb and $^{87}$Rb were (a) $53$ Hz and $ 62$ Hz
on D1 and (b) $95$ Hz and $107$ Hz on D2, respectively. These rates
correspond to a level of random background counts about 2.5 times
lower than the minima of the interference fringes. This implies that
the observed value of visibility $\Upsilon = 0.86$ cannot be
accounted for by random photoelectric coincidences alone. The
additional reduction of visibility may be due to variations in the
idler phases caused by temporal variations in the cloud densities
during data accumulation, while the effects of rf phase noise are
believed to be negligible.

Following Ref. \cite{walls} we calculate the correlation function
$E(\phi_s,\phi_i)$, given by
\begin{equation}
  \frac{C_{si}(\phi_s,\phi_i) -
    C_{si}(\phi_s,\phi_i^\perp) - C_{si}(\phi_s^{\perp},\phi_i) +
    C_{si}(\phi_s^{\perp},\phi_i^{\perp})}
  {C_{si}(\phi_s,\phi_i) +
    C_{si}(\phi_s,\phi_i^\perp) + C_{si}(\phi_s^{\perp},\phi_i) +
    C_{si}(\phi_s^{\perp},\phi_i^{\perp})} \mbox{,}
  \label{eq:Edef}
\end{equation}
where $\phi_{s[i]}^{\perp} = \phi_{s[i]}+\pi$. We note that, by
analogy with polarization correlations, the detected signal [idler]
field $\hat{\psi}_{s[i]}(t,\phi_{s[i]}^{\perp})$ is orthogonal to
$\hat{\psi}_{s[i]}(t,\phi_{s[i]})$, i.e.,
$\left[\hat{\psi}_{s[i]}(t,\phi_{s[i]}),
\hat{\psi}_{s[i]}^{\dag}(t',\phi_{s[i]}^{\perp})\right] = 0$. One
finds that a classical local hidden variable theory yields the Bell
inequality $|S| \le 2$, where $S \equiv
E(\phi_s,\phi_i)-E(\phi_s^{\prime},\phi_i) -
E(\phi_s,\phi_i^{\prime}) - E(\phi_s^{\prime},\phi_i^{\prime})$
\cite{bell}. Using Eq.(\ref{eq:fringes1}), the correlation function
is given by
\begin{equation}
  E(\phi_s,\phi_i) = \Upsilon \cos (\phi_s-\phi_i+\phi_0) \mbox{.}
\end{equation}
Choosing, e.g., the angles $\phi_s=-\phi_0$, $\phi_i = \pi/4$,
$\phi_s^{\prime} = -\phi_0 -\pi/2$, and $\phi_i^{\prime} = 3\pi/4$,
we find the Bell parameter $S = 2\sqrt{2}\Upsilon$.

\begin{table}
\caption{\label{tab:table1} Measured correlation function $E(\phi
_s, \phi _i)$ and $S$ for  $\Delta t = 150$ ns delay between {\it
write} and {\it read} pulses; all the errors are based on the
statistics of the photon counting events.}
\begin{ruledtabular}
\begin{tabular}{ccccc}
$\phi_s$ & $\phi _i $& $E(\phi_s, \phi _i)$  \\
\hline
$0$  & $\pi/4$     &  $0.629  \pm 0.018$   \\
$0$ & $3\pi/4$ &  $-0.591  \pm 0.018$   \\
$-\pi/2$  & $\pi/4$    &  $-0.614  \pm 0.018$   \\
$-\pi/2$  & $3\pi/4$ &  $-0.608 \pm 0.018$   \\
     &    & $S_{exp}=2.44 \pm 0.04$           \\
\end{tabular}
\end{ruledtabular}
\end{table}

Table 1 presents measured values for the correlation function
$E\left({\phi_s},\phi _i\right)$ using the canonical set of angles
$\phi_s,\phi _i$. We find $S_{exp}=2.44 \pm 0.04 \nleq 2$ - a clear
violation of the Bell inequality. This value of $S_{exp}$ is
consistent with the visibility of the fringes $\Upsilon \approx
0.86$ shown in Fig. 3. This agreement supports our observation that
systematic phase drifts are negligible. We emphasize that no active
phase stabilization of any optical frequency field is employed.

In conclusion, we report the first realization of a dual species
matter qubit and its entanglement with a frequency-encoded photonic
qubit. Although we employed two different isotopes, our scheme
should work for chemically different atoms (e.g., rubidium and
cesium) and/or molecules.

This work was supported by NSF, ONR, NASA, Alfred P. Sloan and
Cullen-Peck Foundations. Present addresses: $^{*}$Dipartimento di
Fisica e Matematica, Universit\`{a} dell' Insubria, 22100 Como,
Italy; $^{\star}$Laboratoire Aim\'e Cotton, CNRS-UPR 3321,
B\^atiment 505, Campus Universitaire, 91405 Orsay Cedex, France;
$^{\dagger}$Department of Physics, University of Michigan, Ann
Arbor, Michigan 48109.

\end{document}